\documentclass[a4j]{jpsj3}
\usepackage{bm}
\usepackage{braket}
\usepackage[dvipdfmx]{graphicx}
\usepackage[dvipdfmx]{color}
\usepackage{txfonts}
\usepackage{color,ulem}
\usepackage{amssymb}
\usepackage{diagbox}

\title{Spin-Triplet Superconductivity Induced by All-In-All-Out Spin Fluctuation in Pyrochlore Electronic System}

\author{Akiho \textsc{Sumiyoshiya}$^{1}$ and Tetsuya \textsc{Takimoto}$^{2}$}
\inst{$^{1}$Graduate School of Science and Technology, Niigata University, Niigata 950-2181, Japan \\
$^{2}$Faculty of Engineering, Niigata University, Niigata 950-2181, Japan}

\abst{We study the symmetry of superconductivity around the All-In-All-Out (AIAO) spin ordered phase in a nearly half-filled pyrochlore electronic system. The electronic system is described by the Hubbard model with an anti-symmetric spin-orbit coupling. The AIAO spin ordered state is formed by arranging tetrahedral hedgehog spin-clusters uniformly in the lattice system, where 12 types of spin-clusters are allowed in a unit cell. All spin-cluster susceptibilities are calculated within the random phase approximation with respect to the onsite electron interaction. It is shown that, for the negative spin-orbit coupling coefficient, the AIAO spin-cluster susceptibility diverges at ${\bm q=0}$, while other types of spin-cluster susceptibilities are not developed and are structureless. By calculating the superconducting transition mediated by AIAO spin fluctuation, it is found that a spin-triplet superconductivity appears, while there is no signal of superconducting transition for the spin-singlet one. We suggest that the spin-triplet superconductivity appears around the AIAO spin ordered phase in the pyrochlore lattice.}
\kword{pyrochlore, All-In-All-Out, superconductivity}

\begin{document}

\maketitle

\section{Introduction}

The geometrically frustrated system with strong spin-orbit coupling is an ideal platform for rich electronic phenomena, which will be realized by pyrochlore compounds. Actually, the pyrochlore system shows a variety of electronic states such as metal-insulator transition \cite{MI}, anomalous Hall effect \cite{AHE}, spin liquid state \cite{SL} and Weyl semimetal \cite{Wyle}. 

An interesting series of pyrochlore compounds is $A_{\rm 2}B_{\rm 2}\rm{O_7}$,
where $A$ is $4d-$transition metal or rare earth atom, while $B$ is $5d-$transition metal atom. This system has a cubic crystal structure with the space group $\rm{Fd\bar 3m\ (No. 227, O_h^7)}$. In this crystal structure, both $A$ and $B$ form pyrochlore lattices, which are obtained by corner-sharing tetrahedrons consisting of $A$ and $B$, respectively. The tetrahedron consisting of $A$ includes one oxygen at the center of the tetrahedron, while each $B$ is at the center of O$_6$.

Among $A_{\rm 2}B_{\rm 2}\rm{O_7}$, several experimental measurements have observed the so-called All-In-All-Out (AIAO) type of magnetic structure below $T_{AIAO}$ in pyrochlore iridates $\rm{\it A_2 \rm{Ir}_2 \rm O_7}$ ($A$=Sm, Nd), where all spins of rare earth ions at corners of a tetrahedron orient inward (outward) from the center of the tetrahedron.
This magnetic order will be antiferromagnetic with the ordering wave vector $\bm Q = (0, 0, 0)$ similar to the 120 degrees spin structure of triangular lattice. For $\rm{Nd_2 Ir_2 O_7}$, the origin of the magnetic moment of the AIAO spin-structure has been considered as from Nd$^{3+}$ ion, because the measured value of the magnetic moment of Nd-ion is 2.3$\mu_{\rm B}$ at 0.7 K\cite{AIAONd1} and that of Ir-ion is expected to be 1 $\mu_{\rm B}$ at most\cite{AIAONd2}, where $\mu_B$ is the Bohr magneton.
Interestingly, another compound $\rm{Sm}_2\rm{Ir}_2\rm O_7$ exhibits the metal-insulator transition under the AIAO spin ordered state\cite{AIAO3}, where the Ir-spins form the spin structure, instead of Sm\cite{AIAO2}.

In the same series of pyrochlore compound $\rm{Cd_2Re_2O_7}$, superconductivity has been observed below $T_c\simeq1$K \cite{SC1}. Under the ambient pressure, the symmetry of superconductivity in the compound is considered to be of conventional s-wave, based on the experimental results of specific heat\cite{SC2}, $^{187}\rm{Re}$ nuclear quadrupole resonance (NQR)\cite{SC3}, $^{111}\rm{Cd}$ nuclear magnetic resonance (NMR)\cite{SC4}, and $\mu$SR spectroscopy\cite{SC5}. On the other hand, it has also been considered that, by applying pressure, the symmetry of superconductivity changes from the s-wave to an unconventional spin-triplet p-wave state, 
because the Hebel-Slichter peak disappears increasing pressure and $H_{c2}$ exceeds the value of Pauli limit.\cite{SC6}

In this study, we investigate the symmetry of superconductivity around the AIAO spin ordered phase. At first, we introduce the pyrochlore lattice Hubbard model with an anti-symmetric spin-orbit coupling. Next, we introduce spin-cluster operators for the pyrochlore lattice to calculate spin-cluster susceptibilities within the random phase approximation (RPA) with respect to on-site Coulomb interaction. Finally, we calculate the transition to superconductivity mediated by AIAO spin fluctuation and discuss the symmetry of superconductivity.\cite{pro}

\section{Model and Electronic States}

\begin{figure}[t]
\centering
\includegraphics[scale=0.23]{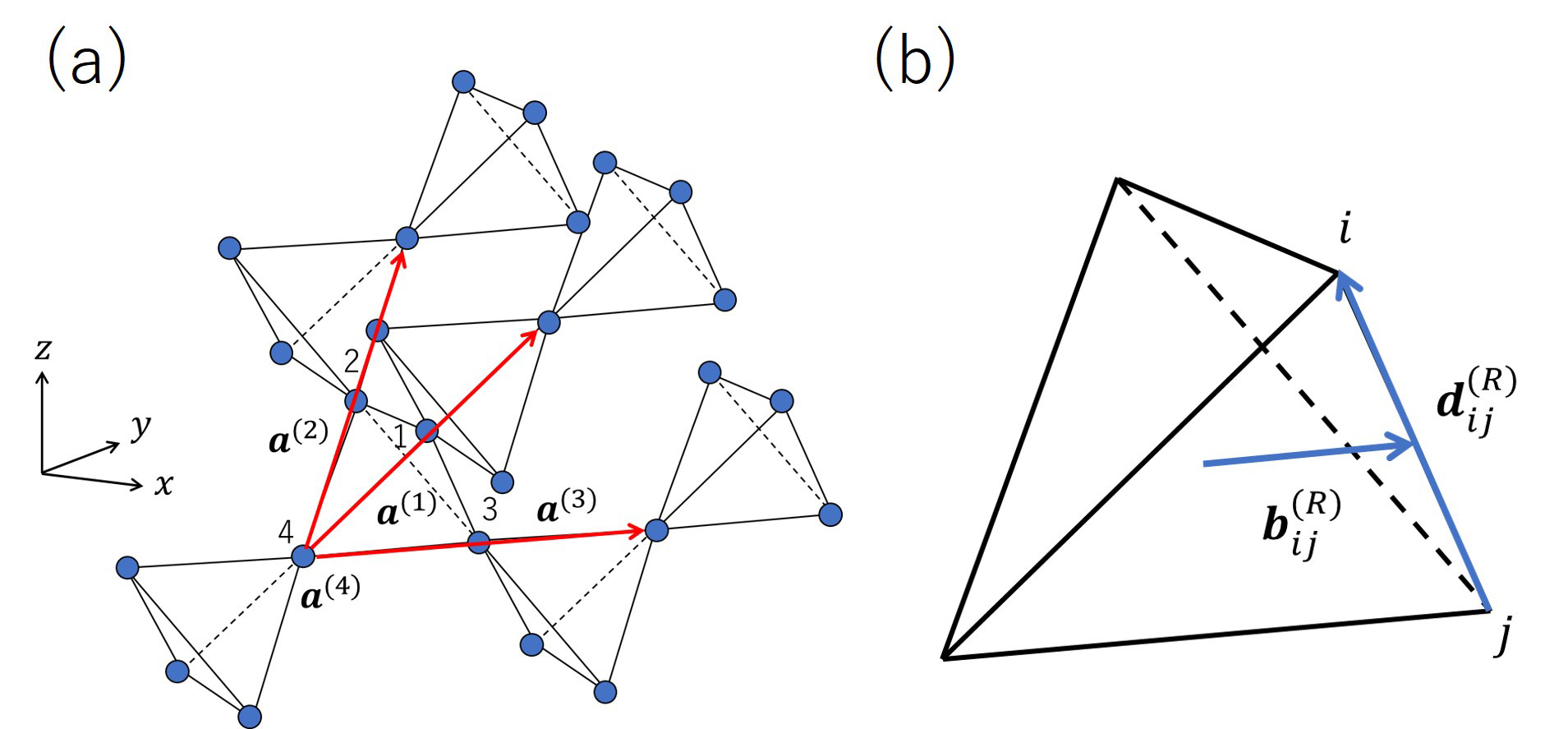}
\caption{(a) The pyrochlore lattice. (b) Definition of $\bm b_{ij}^{(R)}$ and $\bm d_{ij}^{(R)}$ for the right tetrahedron.(Color online)}
\label{f1}
\end{figure}

Now, we construct a microscopic model for the pyrochlore lattice, whose lattice points are given by corners of a tetrahedron, which are shared by the neighboring tetrahedrons, as shown in Fig.\ref{f1}(a). 
Choosing the origin at the center of the tetrahedron in the unit cell, the point group of the system is of $\rm{T_d}$. The representation of the sublattice space $\Gamma_o$ is reduced to $\Gamma_o=\rm{A_1} \oplus \rm{T_2}$ representations of $\rm{T_d}$. The electronic state of the pyrochlore lattice is described by the Hubbard model with an anti-symmetric spin-orbit coupling\cite{STI}, as follows
\begin{align}
H&=H_0 + H_{SO} + H',\\
H_0 
	&=-t\sum_{\bm R}\sum_{<ij>}\sum_\sigma (c_{\bm R i\sigma}^\dag c_{\bm R j\sigma}+c_{\bm R+\bm a^{(j)}-\bm a^{(i)} i\sigma}^\dag c_{\bm R j\sigma}+h.c.), \\
H_{SO}
 &=\sqrt 2 \lambda \sum_{\bm R}\sum_{<ij>}\sum_{\sigma \sigma'}(ic_{\bm R i\sigma}^ \dag \frac{\bm b_{ij}^R \times \bm d_{ij}^R}{|\bm b_{ij}^R \times \bm d_{ij}^R|}\cdot \bm \sigma_{\sigma \sigma'}c_{\bm R j\sigma'}\nonumber\\
&\quad+ ic_{\bm{R}+\bm{a}^{(j)}-\bm a^{(i)}i\sigma}^\dag \frac{\bm b_{ij}^L \times \bm d_{ij}^L}{|\bm b_{ij}^L \times \bm d_{ij}^L|}\cdot \bm \sigma_{\sigma \sigma}c_{\bm R j\sigma'}+h.c.),\\
H'
 &=U\sum_{\bm R}\sum_i c_{\bm R i\uparrow}^\dag c_{\bm R i\uparrow}c_{\bm R i\downarrow}^\dag c_{\bm R i\downarrow},
\label{U}
\end{align}
where $c_{\bm R i\sigma}$ is the annihilation operator for an electron with spin $\sigma$ at sublattice $i$ in a unit cell, whose representative point is given by $\bm R$. $\sum_{<ij>}$ means the sum of the nearest pair of sublattices, and four $\bm a^{(j)}$ are given as $\bm a^{(1)}=\bm a_1$, $\bm a^{(2)}=\bm a_2$, $\bm a^{(3)}=\bm a_3$ and $\bm a^{(4)}=\bm 0$, where $\bm a_1=2a(1,0,1)$, $\bm a_2=2a(0,1,1)$ and $\bm a_3=2a(1,1,0)$ are the primitive translation vectors of the face-centered cubic lattice.
We assume that there is one electronic orbital at each site. 
$H_0$ represents the nearest-neighbor hopping with amplitude $t$, which is the unit of energy in the following.
$H_{SO}$ describes the anti-symmetric spin-orbit coupling satisfying $\rm{T_d}$ point group, where $\lambda$ is a spin-orbit coupling constant and $R\ (L)$ is an index for right (left) tetrahedron. $\bm d_{ij}^{R(L)}$ is the vector from the sublattice $j$ to $i$  and $\bm b_{ij}^{R(L)}$ is the vector from the center of the tetrahedron to the midpoint of the $\bm d_{ij}^{R(L)}$, that $\bm d_{ij}^{R(L)}$ and $\bm b_{ij}^{R(L)}$ are schematically depicted in Fig.\ref{f1}(b). $H'$ is the on-site Coulomb interaction ($U>0$). 
$H'$ includes the intra-orbital Coulomb interaction only ($U>0$), because 
we neglect the electronic orbital degrees of freedom at each site.
\begin{figure}[t]
\centering
\includegraphics[scale=0.34]{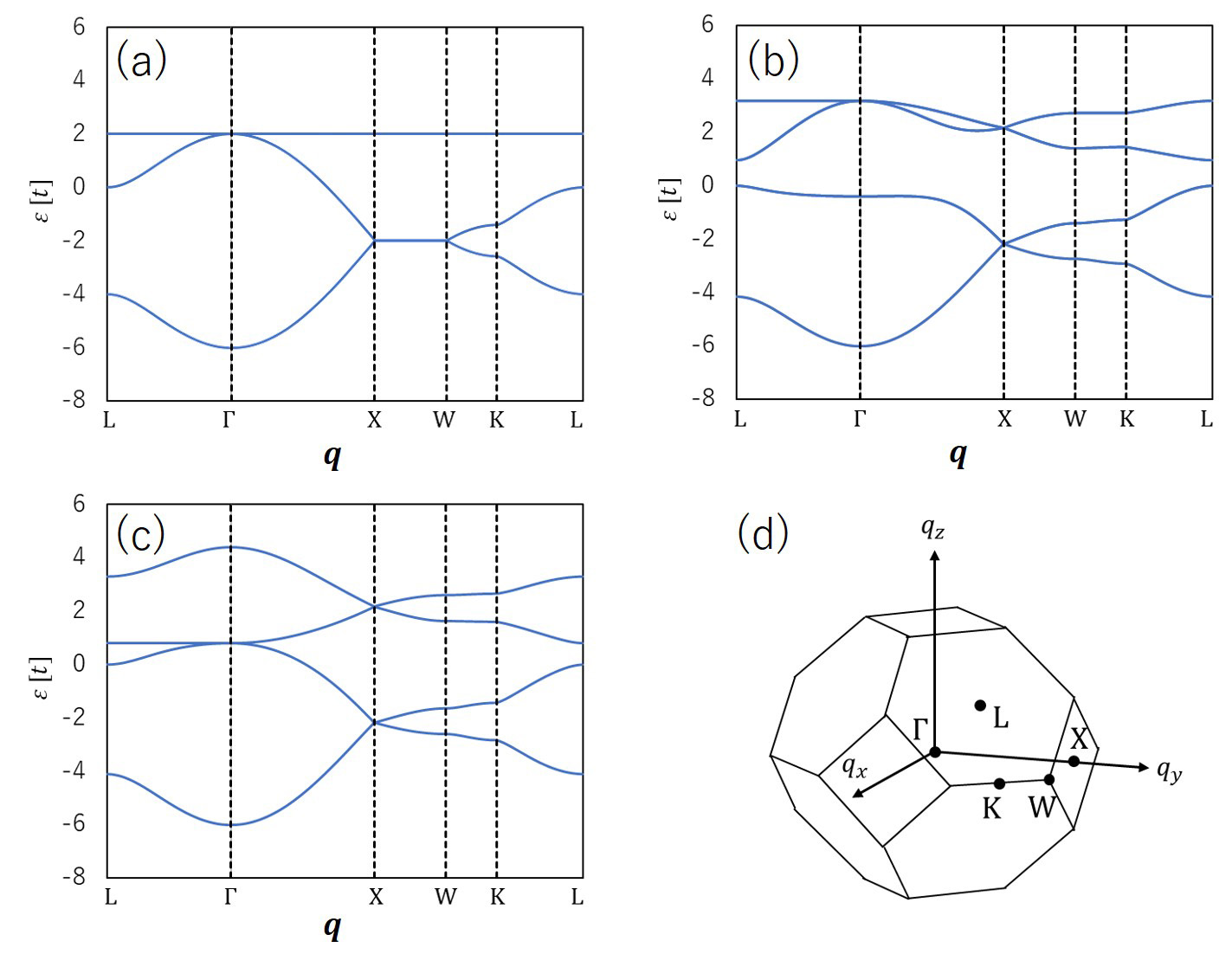}
\caption{(a), (b), (c) Energy bands of $H_0+H_{SO}$ calculated for (a) $\lambda=0$, (b) $\lambda=0.3t$ and (c) $\lambda=-0.3t$, respectively. (d) The first Brillouin zone of a face-centered cubic lattice.(Color online)}
\label{f2}
\end{figure}

Figures \ref{f2}(a), (b), and (c) show the energy bands of $H_0 + H_{SO}$ calculated for $\lambda = 0, 0.3t$ and $-0.3t$, respectively. Without spin-orbit interaction, there is a four-fold degenerated flat band at $\epsilon(\bm k)=2t$.
For $\lambda<0$, there is no energy gap between neighboring bands and the electronic structure is semi-metallic at half-filling.
On the other hand, for $\lambda>0$ the energy gap opens between 2nd- and 3rd- bands to be a topological insulator for the half-filling.\cite{STI}

In most of the remaining, we focus our attention on $\lambda<0$, because the AIAO spin order appears for $\lambda<0$ only. 
On the other hand, by changing the sign of $\lambda$, the type of spin-cluster 
order is different from the AIAO-type, as shown in the previous study based on the pyrochlore spin system with the corresponding Dzyaloshinskii-Moriya interaction.\cite{LAM}

\section{Spin-cluster Susceptibility}
In order to find the dominant low-energy spin fluctuation which contributes to the superconducting transition in the pyrochlore lattice, we calculate spin cluster susceptibilities. Since there are four sublattices in the unit cell, first of all, we introduce spin-cluster operators of the pyrochlore lattice. Four one-electronic orbital states in a tetrahedron are classified into the irreducible representations $\Gamma_o = \rm A_1 \oplus \rm T_2$. 
These bases will be defined as follows,
\begin{align}
\phi_{\rm A_1} &= \frac{1}{2}(f_1+f_2+f_3+f_4),\\
\phi_{\rm T_2\xi} &= \frac{1}{2}(f_1-f_2-f_3+f_4),\\
\phi_{\rm T_2\eta} &= \frac{1}{2}(-f_1+f_2-f_3+f_4),\\
\phi_{\rm T_2\zeta} &= \frac{1}{2}(-f_1-f_2+f_3+f_4),
\end{align}
where $f_i\ (i=1, 2, 3,4)$ is the basis in the sublattice space. Noting that the representation of the spin operators at every sublattice $\Gamma_{\rm s}$ is classified into $D^{(1)} \downarrow \rm{T_d}=\rm T_1$, the representation of spin-cluster is reduced to $\Gamma_s \otimes \rm{\Gamma_o} =\rm A_2 \oplus \rm E \oplus 2\rm T_1 \oplus \rm T_2$ including 12 components. The bases of this product representation are obtained by using the Clebsch-Gordan coefficients\cite{CG}. In particular, the AIAO spin cluster belonging to $\rm A_2$ representation is described as follows:
\begin{align}
\hat S_{A_2}(\bm R)
	&= -\frac{1}{2\sqrt 3}\left(\hat S_{\bm R1}^x-\hat S_{\bm R2}^x-\hat S_{\bm R3}^x+\hat S_{\bm R4}^x-\hat S_{\bm R1}^y+\hat S_{\bm R2}^y\right.\nonumber\\
&\left.\quad-\hat S_{\bm R3}^y+\hat S_{\bm R4}^y-\hat S_{\bm R1}^z-\hat S_{\bm R2}^z+\hat S_{\bm R3}^z+\hat S_{\bm R4}^z\right),
\end{align}
where $\hat S_{\Gamma\gamma}(\bm R)$ is a spin-cluster operator of $\gamma$-component in $\Gamma$ representation whose representative point is given by $\bm R$. $\hat S_{\bm Ri}^\alpha=\frac{1}{2}\sum_{\sigma,\sigma'}\sigma_{\sigma,\sigma'}^\alpha c_{\bm Ri\sigma}^\dag c_{\bm Ri\sigma'}\ (\alpha=x,y,z)$ is $\alpha$-component of the spin operator at sublattice-$i$ in the unit cell at $\bm R$. Other spin-cluster operators are also described as linear combinations of $\hat S_{\bm Ri}^{\alpha}$ (Appendix A).

In the following, we formulate the spin-cluster susceptibility. 
At first, the spin-cluster susceptibility is described according to the Kubo formula, as follows
\begin{align}
\chi_{\Gamma\gamma,\Gamma'\gamma'}(\bm q, i\Omega_m)
	=\frac{1}{N}\int_0^\beta d\tau e^{i\Omega_m \tau}\braket{T_\tau [\hat S_{\Gamma\gamma}(\bm q, \tau)\hat S_{\Gamma'\gamma'}(-\bm q, 0)]},
\end{align}
where $\braket \cdots$ is the thermal average with the total Hamiltonian $\hat{H}$ and 
$A(\tau)=e^{(\hat{H}-\mu\hat{N})\tau}A\:e^{-(\hat{H}-\mu\hat{N})\tau}$, where $\mu$ is the chemical potential. 
Then, noting that all spin(charge)-cluster operators $\hat{S}_{\Gamma\gamma}$ ($\hat \rho_{\Gamma\gamma}$) are given by linear combinations of $\hat{S}^{\alpha}_{\bm{R}i}$ ($\hat S_{\bm Ri}^c=\frac{1}{2}\sum_\sigma c_{\bm Ri\sigma}^\dag c_{\bm Ri\sigma}$)
which are given by linear combinations of four particle-hole pairs at each sublattice, 
these transformations are described as
  \begin{align}
\begin{pmatrix}
	\hat S_i^c\\
	\hat S_i^z\\
	\hat S_i^x\\
	\hat S_i^y\\
\end{pmatrix}
=\hat W_{1i}
\begin{pmatrix}
	S_i^{\uparrow\uparrow}\\
	S_i^{\downarrow\downarrow}\\
	S_i^{\uparrow\downarrow}\\
	S_i^{\downarrow\uparrow}\\
\end{pmatrix},\qquad
\begin{pmatrix}
	\rho_{\rm{A_2}}\\
	\vdots\\
	\rho_{\rm{T_2\zeta}}\\
	S_{\rm{A_2}}\\
	\vdots\\
	S_{\rm{T_2\zeta}}
\end{pmatrix}
=\hat W_2
\begin{pmatrix}
	\hat S_1^c\\
	\vdots\\
	\hat S_4^c\\
	\hat S_1^z\\
	\vdots\\
	\hat S_4^y
\end{pmatrix}.
\label{uniW}
\end{align}
Here, $\hat{W}_{1i}$ and $\hat{W}_2$ are corresponding $4\times 4$ and $16\times 16$ unitary matrices, respectively. 
Therefore, the spin-cluster susceptibility $\chi_{\Gamma\gamma,\Gamma'\gamma'}(\bm q, i\Omega_m)$ is given by the linear combination of 
\begin{align}
\tilde{\chi}& _{i\sigma_1\sigma_2, j\sigma_3\sigma_4}(\bm q,i\Omega_m)
	=\frac{1}{N}\int_0^\beta d\tau e^{i\Omega_m\tau}e^{i\bm q\cdot (\bm R_i-\bm R_j)}\nonumber\\
&\hspace{10mm}\times\sum_{\bm k,\bm k'}\braket{T_\tau[c_{\bm ki\sigma_1}^\dag(\tau)c_{\bm k+\bm qi\sigma_2}(\tau)c_{\bm k'j\sigma_4}^\dag c_{\bm k'-\bm qj\sigma_3}]},
\label{sosus}
\end{align}
due to a unitary matrix $\hat{W}_2\hat{W}_1$, where $\hat{W}_1$ is a $16\times 16$ matrix, whose $i$-th $4\times 4$ diagonal block matrix is given by $\hat{W}_{1i}$ and all off-diagonal ones are zero. 
Namely, the spin-cluster susceptibility matrix $\hat{\chi}(\bm q, i\Omega_m)$ is obtained from $\hat{\tilde{\chi}}(\bm q, i\Omega_m)$ through the unitary transformation, as follows
\begin{align}
\hat{\chi}(\bm q, i\Omega_m)
 =\hat{W}_2\hat{W}_1\hat{\tilde{\chi}}(\bm q, i\Omega_m)\hat{W}_1^{\dagger}\hat{W}_2^{\dagger}.
\end{align}

Applying the diagrammatic technique, the susceptibility matrix $\hat{\tilde{\chi}}(\bm q, i\Omega_m)$ 
will be described as
\begin{align}
\hat{\tilde{\chi}}(\bm q, i\Omega_m)
 &=\hat{\bar{\tilde{\chi}}}(\bm q, i\Omega_m)
 +\hat{\bar{\tilde{\chi}}}(\bm q, i\Omega_m)\hat{\tilde{U}}\hat{\tilde{\chi}}(\bm q, i\Omega_m)\nonumber\\
 &=\hat{\bar{\tilde{\chi}}}(\bm q, i\Omega_m)[1-\hat{\tilde{U}}\hat{\bar{\tilde{\chi}}}(\bm q, i\Omega_m)]^{-1},
\end{align}
where $\hat{\bar{\tilde{\chi}}}(\bm q, i\Omega_m)$ is the irreducible susceptibility matrix of $\hat{\tilde \chi}(\bm q, i\Omega_m)$, 
whose matrix element is defined by a set of diagrams that cannot be separated into two pieces by removing an interaction vertex. Further, $\hat{\tilde{U}}$ is given by a $16\times 16$ matrix, 
whose off-diagonal $4\times4$ matrices are zero and diagonal four $4\times 4$ matrices are given by 
\begin{eqnarray}
 \hat{\tilde{U}}_i=
 \left(
  \begin{array}{cccc}
   0 & -U & 0 & 0\\
   -U & 0 & 0 & 0\\
   0 & 0 & U & 0\\
   0 & 0 & 0 & U\\
  \end{array}
  \right)
\end{eqnarray}
with the basis $\{\uparrow\uparrow, \downarrow\downarrow, \uparrow\downarrow,\downarrow\uparrow\}$ of spins $(\sigma \sigma')$ of the particle-hole pair.
Then, using the unitary transformation given above, the spin-cluster susceptibility matrix is given by
\begin{align}
\hat{\chi}(\bm q, i\Omega_m)
 &=\hat{\bar{\chi}}(\bm q, i\Omega_m)[1-2\hat{U}\hat{\bar{\chi}}(\bm q, i\Omega_m)]^{-1},
\end{align}
where $\hat{\bar{\chi}}(\bm q, i\Omega_m)$ is the irreducible spin-cluster susceptibility matrix of $\hat \chi(\bm q, i\Omega_m)$ and 
\begin{align}
\hat U
	=\begin{pmatrix}
	\hat U_\rho \\
	& \hat U_S
\end{pmatrix}
\end{align}
with $4\times 4$ and $12\times 12$ matrices $\hat U_\rho=-U\:\hat{1}_4$ and $\hat U_S=U\:\hat{1}_{12}$. 

In order to calculate the spin-cluster susceptibility matrix, we apply the random phase approximation (RPA). 
In the present formalism, the RPA spin-cluster susceptibility matrix is obtained by 
choosing $\hat{\bar{\chi}}(\bm q, i\Omega_m)$ as the noninteracting one, 
\begin{align}
\bar{\chi}_{\Gamma\gamma,\Gamma'\gamma'}^0(\bm q, i\Omega_m)
	=\frac{1}{N}\int_0^\beta d\tau e^{i\Omega_m \tau}\braket{T_\tau [\hat S_{I\Gamma\gamma}(\bm q, \tau)\hat S_{I\Gamma'\gamma'}(-\bm q, 0)]}_0,
\end{align}
where $\braket \cdots _0$ is the thermal average with the noninteracting Hamiltonian $\hat{H}_0+\hat{H}_{SO}$ and 
$A_I(\tau)=e^{(\hat{H}_0+\hat{H}_{SO}-\mu\hat{N})\tau}A\:e^{-(\hat{H}_0+\hat{H}_{SO}-\mu\hat{N})\tau}$. 
In the following, we call the diagonal components of $\hat{\chi}(\bm q, i\Omega_m)$ as 
$\chi_{\Gamma\gamma}(\bm q, i\Omega_m)$. Especially, the $\Gamma=$A$_2$ spin-cluster susceptibility within RPA 
is written as the scalar form
\begin{align}
 \chi_{\rm{A_2}}(\bm q, i\Omega_m)=\frac{{\bar \chi}_{\rm{A_2}}(\bm q, i\Omega_m)}{1-2U{\bar \chi}_{\rm{A_2}}(\bm q, i\Omega_m)}.
\end{align}

\begin{figure}[t]
\begin{center}
\includegraphics[scale=0.55]{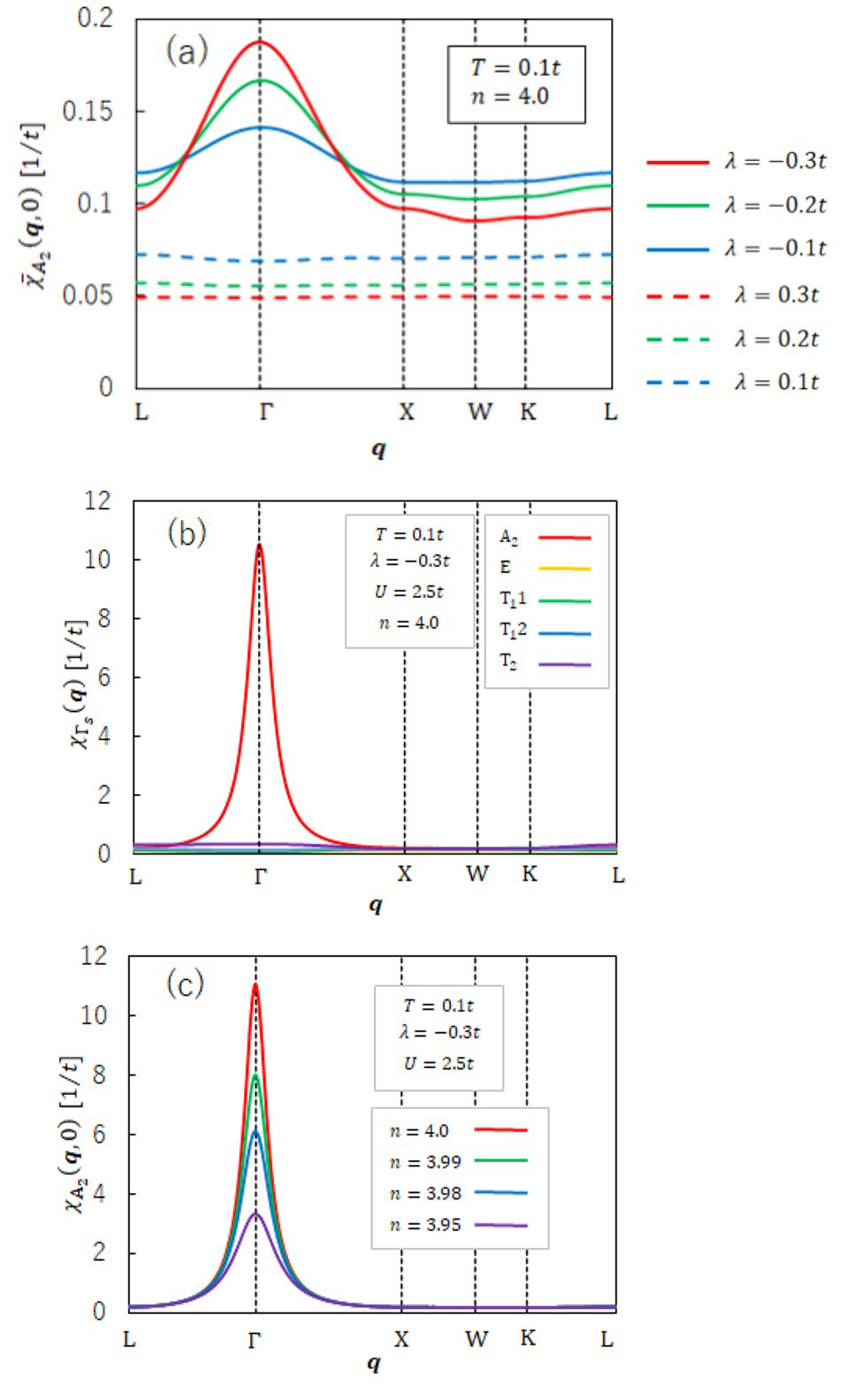}
\caption{(a) $\bm q$-dependences of irreducible susceptibility of $\rm{A_2}$ representation $\bar \chi_{\rm{A_2}}(\bm q)$ for various values of $\lambda$. (b) $\bm q$-dependences of RPA susceptibility of all spin-clusters with $T=0.1t$, $\lambda=-0.3t$, $U=2.5t$ and $n=4.0$. For multi-dimensional representations, the maximum eigenvalues of $\hat \chi(\bm q)$ are plotted. (c) $\bm q$-dependences of RPA susceptibility of $\rm{A_2}$ representation for various filling ($n<4.0$).(Color online)}
\label{f3}
\end{center}
\end{figure}

We show that the AIAO spin-cluster fluctuation develops dominantly among all spin-cluster fluctuations for $\lambda<0$. Fig.\ref{f3}(a) shows $\bm q$-dependences of irreducible susceptibility $\bar \chi_{\rm{A_2}}(\bm q, 0)$ for various values of $\lambda$ at the half-filling. The susceptibility for $\lambda<0$ is larger than $\lambda>0$ over the entire $\bm q$-space and develops at $\bm q=0$ for $\lambda<0$ only. It is suggested that the AIAO spin fluctuation develops only for $\lambda<0$. Due to this result, we consider only $\lambda<0$ in the following. 

Figure \ref{f3}(b) shows $\bm q$-dependences of the static RPA spin-cluster susceptibilities $\chi_{\Gamma\gamma}(\bm q, 0)$ for half-filling ($n=4.0$). The RPA susceptibility $\chi_{\rm{A_2}}(\bm q, 0)$ shows a divergent behavior at $\Gamma$-point, which is the ordering wave vector of the AIAO spin order.  Therefore, the AIAO spin-cluster fluctuation becomes dominant among not only all types of spin-clusters but also ordering wave vectors for $\lambda<0$. 

This feature is robust against changing the electron density around the half-filling, as shown in Fig. \ref{f3}(c). 
Therefore, the development of AIAO spin fluctuation is kept even in the metallic case 
for the negative value of $\lambda$, to discuss the superconducting transition mediated by the AIAO 
spin fluctuation. 

\section{Superconducting Transition Induced by AIAO Spin Fluctuation}
In this section, we study what kind of superconductivity appears in the system close to 
the AIAO spin ordered phase, where the AIAO spin fluctuation develops only among all spin-cluster fluctuations.
At first, we examine the possible type of the Cooper pair in the pyrochlore electronic system.
Since the unit cell of pyrochlore lattice has four sublattices, we need to consider not only wave-vector and spin spaces but also sublattice space to describe the type of the Cooper pair (Table \ref{t1}).\cite{SCcal} 
Therefore, the representation of superconductivity $\Gamma_{\rm SC}$ is given as $\Gamma_{\rm SC}=\Gamma_{\bm k}\otimes \tilde \Gamma_s\otimes\tilde{\Gamma}_o$, where $\Gamma_{\bm k}$, $\tilde \Gamma_s$ ($= (D^{(0)}\oplus D^{(1)}) \downarrow \rm{T_d}=\rm A_1\oplus \rm T_1$), and $\tilde{\Gamma}_o$ ($= \Gamma_o \otimes \Gamma_o$) are representations for the wave-vector space, the spin space, and the sublattice space of the gap function, respectively.
Among the four cases shown in Table \ref{t1}, it will be sufficient to consider the sublattice-symmetric pairs only in the metallic 
case of $n\approx 4.0$ for $\lambda<0$, since there is only one band that crosses the Fermi level. In this case, $\tilde{\Gamma}_o$ is approximated to symmetric representation in the sublattice space $[\Gamma_o\otimes\Gamma_o]=2\rm{A_1}\oplus\rm E \oplus 2\rm{T_2}$.

In order to study the superconducting transition in the pyrochlore electronic system 
close to the AIAO spin ordered phase, we will give the corresponding linearized gap equation. 
The temperature Green's function and the anomalous Green's function are defined as follows,
\begin{align}
&G_{\sigma\sigma'}^{ij}(\bm k, \tau)
	=-\braket{T_\tau[c_{\bm ki\sigma}(\tau)c_{\bm kj\sigma'}^\dag (0)]},\\
&F_{\sigma\sigma'}^{ij}(\bm k, \tau)
	=-\braket{T_\tau[c_{\bm ki\sigma}(\tau)c_{-\bm kj\sigma'}(0)]}.
\end{align}
According to the weak coupling theory for superconductivity, we obtain the Dyson-Gorkov equation,
\begin{align}
&\hat G(\bm k, i\omega_n)
	=\hat G^{(0)}(\bm k, i\omega_m) + \hat G^{(0)}(\bm k, i\omega_m)\hat \Delta(\bm k)\hat F(\bm k, i\omega_m),\\
&\hat F(\bm k, i\omega_m)
	=-\hat G^{(0)}(\bm k, i\omega_m)\hat \Delta(\bm k)\hat G(-\bm k, -i\omega_m),
\label{F}
\end{align}
where $\omega_m$ is the fermionic Matsubara frequency. $\hat G^{(0)}$ is the non-interacting temperature Green's function with Hamiltonian $H_0+H_{SO}$, given by
\begin{align}
G_{\sigma,\sigma'}^{(0)ij}(\bm k, i\omega_m)
	=\sum_n (U_{\bm k})_{i\sigma,n}(U_{\bm k})_{j\sigma',n}^*\frac{1}{i\omega_m-\xi_{\bm kn}},
\end{align}
where $\xi_{\bm kn}=\epsilon_{\bm kn}-\mu$ and $\epsilon_{\bm k n}$ is the $n$-th energy band of Hamiltonian $H_0+H_{SO}$. $\hat U_{\bm k}$ is the unitary matrix from the sublattice-spin basis $\{l\sigma\}$ to the band basis $\{n\}$.
The linearized gap equation is given by 
\begin{align}
\Delta_{\sigma_1 \sigma_2}^{l_1l_2}(\bm k)
	=&\sum_{\bm k'}\sum_{l_5,l_6}\sum_{\sigma_3 \sigma_4, \sigma_5,\sigma_6}\sum_{n_5,n_6}V_{\sigma_1 \sigma_2, \sigma_3 \sigma_4}^{l_1 l_2}(\bm k,\bm k')\nonumber\\
	&\times(U_{\bm k'})_{l_1\sigma_3,n_5}(U_{\bm k'})_{l_5\sigma_5,n_5}^*(U_{-\bm k'})_{l_2\sigma_4,n_6}
	 (U_{-\bm k'})_{l_6\sigma_6,n_6}^*\nonumber\\
	&\times\frac{f(\xi_{\bm k'n_5})+f(\xi_{-\bm k'n_6})}{\xi_{\bm k'n_5}+\xi_{-\bm k'n_6}}
	 \Delta_{\sigma_5\sigma_6}^{l_5l_6}(\bm k),
\end{align}
with the pairing interaction mediated by the AIAO fluctuation $\chi_{\rm{A_2}}(\bm k-\bm k')$
\begin{align}
&V_{\sigma_1\sigma_2\sigma_3\sigma_4}^{l_1l_2}(\bm k,\bm k')\nonumber\\
	&=\frac{U}{N}\delta_{l_1,l_2}\delta_{\sigma_2,\bar\sigma_1}\delta_{\sigma_1,\sigma_3}\delta_{\sigma_2,\sigma_4}-\frac{U^2}{N}e^{-i(\bm k-\bm k')\cdot(\bm R_{l_1}-\bm R_{l_2})}\sigma_1\sigma_2\sigma_3\sigma_4\nonumber\\
	&\times\sum_{\alpha,\alpha'}(W_{1}^{*})_{l_1\alpha,l_1\bar{\sigma_1}\bar{\sigma_3}}(W_1)_{l_2\alpha',l_2\bar{\sigma_4}\bar{\sigma_2}} (W_{2}^{*})_{{\rm A_2}, l_1\alpha}(W_{2})_{{\rm A_2},\it{l}_2\alpha'}\chi_{\rm{A_2}}(\bm k-\bm k'),
\end{align}
where $f(\xi_{\bm kn})$ is Fermi distribution function of $\xi_{\bm kn}$. 
The paring interaction is obtained as follows; (1) the expression of the fluctuation-exchange term is given first within the second-order perturbation of $\hat F(\bm k, i\omega_m)$ with respect to $U$, (2) expressions of noninteracting fluctuations are replaced by those of RPA ones, and (3) $\chi_{\rm{A_2}}(\bm k-\bm k')$ is remained only among all fluctuations. Detailed derivations are summarized in Appendix B.
This equation is the same form as the eigenvalue problem with the eigenvector $\{\hat \Delta(\bm k)\}$. 
The superconducting transition is obtained at $U^*$ for fixed temperature, 
where the maximum eigenvalue reaches the unity by increasing the interaction constant $U$. 

\begin{table}[t]
	\caption{ Possible types of the Cooper pair.}
    \label{t1}
	\centering
	\begin{tabular}{cccc}
	&Parity & Spin & Sublattice \\ \hline\hline
	(1) & Even & Singlet & Symmetric\\
	(2) & Even & Triplet & Anti-symmetric\\	
	(3) & Odd & Singlet & Anti-symmetric\\
	(4) & Odd & Triplet & Symmetric \\ \hline
    \end{tabular}
\end{table}

\begin{figure}[t]
\centering
\includegraphics[scale=0.7]{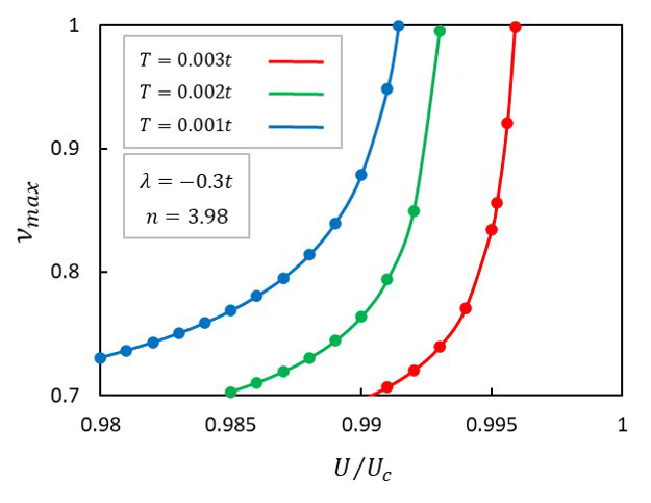}
\caption{The $U$-dependences of the maximum eigenvalue of the linearized gap equation for spin-triplet superconductivity. $U_c$ at $T=0.001t, 0.002t$, and $0.003t$ are calculated to be $2.478t, 2.476t,$ and $2.468t$, respectively.(Color online)}
\label{f5}
\end{figure} 
The maximum eigenvalue $\nu_{max}$ of the linearized gap equation is calculated numerically with the lattice size of $32\times32\times32$ to show a spin-triplet superconducting transition.
Both maximum eigenvalues of the spin-singlet and spin-triplet superconducting transitions are calculated separately, 
though the spin-singlet and spin-triplet superconducting states will be mixed in the  T$_{\rm d}$ point group. 
Fig.\ref{f5} shows the $U$-dependence of the maximum eigenvalue $\nu_{max}$ for spin-triplet superconductivity, where $U_c$ is the critical interaction constant to the AIAO spin ordered state which is satisfied 
by $2U_c\bar \chi_{\rm A_2}(\bm 0) = 1$. It is shown that increasing the value of $U$, the spin-triplet superconducting 
transition takes place in the $U$ region close to $U_c$. 
On the other hand, no spin-singlet superconducting transition takes place, because the maximum eigenvalue is at most 
0.1. Thus, it is shown that spin-triplet superconductivity appears around the AIAO spin ordered phase of the pyrochlore electronic system.

Next, it will be necessary to comment on the irreducible representation $\Gamma_{\bm k}^{irr}$ for ($\alpha, \gamma$)-component of $\bm d$-vector of spin-triplet superconductivity, where $\alpha$ and $\gamma$ are bases in the spin space and the sublattice space, respectively.
In the spin-triplet case neglecting the sublattice-antisymmetric components, the $\bm d$-vector has 30 components at a wave-vector $\bm k$, where the symmetric representation in the sublattice space is given by $2\rm A_1 \oplus \rm E \oplus 2\rm T_2$ including 10 bases.
When spin-triplet superconductivity belonging to the irreducible representation $\tilde \Gamma_{SC}^{irr}$ appears in the system, the relevant irreducible representation of $\bm k$-dependence of $\bm d$-vector component $\Gamma_{\bm k}^{irr}(\tilde \Gamma_o^{irr})$ satisfies the relations $\Gamma_{\bm k}^{irr}(\tilde{\Gamma}_o^{irr})\otimes \tilde{\Gamma}_s^{irr}\otimes \tilde{\Gamma}_o^{irr}=\tilde \Gamma_{SC}$ with $\tilde \Gamma_{SC}^{irr}\in \tilde \Gamma_{SC}$, where $\tilde{\Gamma}_s^{irr}=\rm T_1$ for spin-triplet superconductivity and $\tilde{\Gamma}_o^{irr}$ is an irreducible representation among $\tilde \Gamma_o=2\rm A_1 \oplus \rm E \oplus \rm T_2$.
Further, $\Gamma_{\bm k}^{irr}(\tilde \Gamma_o^{irr})$ is the irreducible representation of $\bm k$-dependence of $\bm d$-vector component, when the irreducible representation in the sublattice symmetric space is of $\tilde{\Gamma}_o^{irr}$.
Now, $\Gamma_{\bm k}^{irr}(\tilde{\Gamma}_o^{irr})$ is obtained, as follows. 
Suppose that for $\tilde \Gamma_o^{irr}=\rm{A_1}$, the irreducible representation of $\bm k$-dependence of $d^{\alpha}_{\rm A_1}({\bm k})$ is of $\Gamma_{\bm k}^{irr}$, $\tilde \Gamma_{SC}=\Gamma_{\bm k}^{irr}\otimes \rm T_1$ with $\tilde \Gamma_{SC}^{irr}\in \tilde \Gamma_{SC}$ is satisfied. 
Simultaniously, the irreducible representation of $\bm k$-dependence of $d^{\alpha}_{\Gamma_o^{irr}}({\bm k})$ should be included in the reducible representation of $\Gamma_{\bm k}^{irr}\otimes \tilde \Gamma_o^{irr}$.

In order to identify the irreducible representation of spin-triplet superconductivity $\tilde \Gamma_{SC}^{irr}$ taking into account these, we apply the projection operator $P^{(\tilde \Gamma_{SC}^{irr})}$ to the eigenvector obtained by the numerical calculation
\begin{align}
P^{(\tilde \Gamma_{SC}^{irr})}&=\frac{d_{\tilde \Gamma_{SC}^{irr}}}{g}\sum_R \chi^{(\tilde \Gamma_{SC}^{irr})}(R)^* R,
\end{align}
where $d_{\tilde \Gamma_{SC}^{irr}}$ is the dimension of $\tilde \Gamma_{SC}^{irr}$, $g$ is the order of $\rm{T_d}$ point group, $R\in\rm{T_d}$ is the symmetrical operator and $\chi^{(\tilde \Gamma_{SC}^{irr})}(R)$ is the character of  $R$ for $\tilde \Gamma_{SC}^{irr}$. 
The symmetrical operator $R$ will be applied to every component of $\bm d$-vector, as follows
\begin{align}
Rd_{\tilde \Gamma_o^{irr}\gamma}^{\alpha}(\bm k)
	&=\sum_{\gamma'}\sum_{\alpha'}D_{\gamma'\gamma}^{(\tilde \Gamma_o^{irr})}(R)D_{\alpha'\alpha}^{(\rm T_1)}(R)d_{\tilde \Gamma_o\gamma'}^{\alpha'}(\hat D^{(\rm T_2)}(R^{-1})\bm k)
\label{R},
\end{align}
where $d_{\tilde \Gamma_o^{irr}\gamma}^{\alpha}(\bm k)$ is the $(\alpha,\gamma)$-component of $\bm d$-vector with $\alpha$-basis of the irreducible representation in spin space $\tilde \Gamma_s^{irr} =\rm T_1$ and $\gamma$-basis of the irreducible representation in sublattice space $\tilde \Gamma_o^{irr}$. $\hat D^{(\Gamma)}(R)$ is the representation matrix of $R$ for representation $\Gamma$.

Especially, for $\tilde \Gamma_o^{irr}=\rm A_1$, because the representation matrix of any symmetrical operator $D^{(\tilde \Gamma_o^{irr})}(R)$ is 1, $d_{A_1}^{\alpha}(\bm k)$ is transformed by $R$ as
\begin{align}
Rd_{A_1}^{\alpha}(\bm k)
	&=\sum_{\alpha'}D_{\alpha'\alpha}^{(\rm T_1)}(R)d_{A_1}^{\alpha'}(\hat D^{(\rm T_2)}(R^{-1})\bm k)
\label{R},
\end{align}
which is similar to the single band case.
Here, we extract a 3-dimensional vector $\tilde{\bm d}_{\rm A_1}(\bm k)=(d_{\rm A_1}^x(\bm k), d_{\rm A_1}^y(\bm k),d_{\rm A_1}^z(\bm k))$.
Now, for spin-triplet superconductibity belonging to $\tilde \Gamma_{SC}^{irr}$ irreducible representation, the 30 dimentional $\bm d$-vector $\bm d(\bm k)$ belongs to the $\tilde \Gamma_{SC}^{irr}$, so that the relation $P^{(\tilde \Gamma_{SC}^{irr})}\tilde{\bm d}_{\rm A_1}(\bm k)=\tilde{\bm d}_{\rm A_1}(\bm k)$ will be satisfied.
On the other hand, if the irreducible representation $\Gamma$ is different from $\tilde \Gamma_{SC}^{irr}$,  $P^{(\Gamma)}\tilde{\bm d}_{\rm A_1}(\bm k)=0$.
Therefore, we can find the irreducible representation of spin-triplet superconductivity $\tilde \Gamma_{SC}^{irr}$ by calculating the ratio of $P^{(\Gamma)}\tilde{\bm d}_{\rm A_1}(\bm k)$ to $\tilde{\bm d}_{\rm A_1}(\bm k)$ for all irreducible representation $\Gamma$.

Table \ref{t2} shows the average for the ratio of $P^{(\Gamma)}\tilde{\bm d}_{\rm A_1}(\bm k)$ to $\tilde{\bm d}_{\rm A_1}(\bm k)$ over momenta close to the Fermi surface. 
The result of $\Gamma=\rm T_1$ yields a value close to 1, while the values of the ratio for the other representations are close to 0. 
Thus, we propose that the irreducible representation of spin-triplet superconductivity around the AIAO spin ordered phase is of $\rm T_1$ of $\rm{T_d}$ point group.

\begin{table}[t]
	\caption{The ratio of $P^{(\Gamma)}\tilde{\bm d}_{\rm A_1}(\bm k)$ to $\tilde{\bm d}_{\rm A_1}(\bm k)$.}
    \label{t3}
	\centering
	\begin{tabular}{cc}
	$\Gamma$ & $P^{(\Gamma)}\tilde{\bm d}_{\rm A_1}(\bm k)$/$\tilde{\bm d}_{\rm A_1}(\bm k)$ \\ \hline\hline
	$\rm A_1$ & -0.0002536\\
	$\rm A_2$ & -0.004040\\	
	$\rm E$ & -0.0008791\\
	$\rm T_1$ & 1.015\\
	$\rm T_2$ & -0.01020 \\ \hline
    \end{tabular}
\label{t2}
\end{table}

\section{Source of Spin-Triplet Superconductivity}
In this section, we will examine the source of spin-triplet superconductivity in the pyrochlore electronic system close to the AIAO spin ordered phase. Since the AIAO spin ordered state will be a three-dimensional analogue of the 120 degrees spin structure stabilized by the antiferromagnetic spin couplings in the triangular lattice, spin-singlet superconductivity will be naively anticipated rather than the spin-triplet one. 
In the following, we will calculate the expectation value of $\hat{S}_{A_2}$ of one tetrahedron by the All-Out spin state and examine the contributions from spin-singlet and spin-triplet spin-pairs.

The All-Out spin state of a tetrahedron will be described by the ground state of a localized spin Hamiltonian {$H_{AO}$ with an effective magnetic field, as follows:
\begin{align}
H_{AO} = -h\sum_l \bm e_l\cdot \hat{\bm S}_l=-2h\hat{S}_{A_2},
\end{align}
with the unit vector ${\bm e}_l$ directed from the center of tetrahedron toward $l$-site describing
the direction of the local effective magnetic field for the All-Out spin state
\begin{align}
\bm e_1 = \frac{1}{\sqrt 3}(-1, 1, 1), &\qquad	\bm e_2 = \frac{1}{\sqrt 3}(1, -1, 1),\nonumber\\
\bm e_3 = \frac{1}{\sqrt 3}(1, 1, -1), &\qquad	\bm e_4 = \frac{1}{\sqrt 3}(-1, -1, -1),
\end{align}
where $\hat S_{\rm A_2}$ is nothing but the All-Out spin operator. Since the Hamiltonian $H_{AO}$ is a block-diagonalized matrix in the basis $\{|l,\uparrow\rangle,
|l,\downarrow\rangle\} (l=1,2,3,4)$, the ground state is easily obtained by diagonalizing the $2\times 2$ block matrix for $l-$site in terms of a unitary matrix $\hat{U}'_l$. 
Since the ground state is the All-Out state, the All-Out spin state is given by 
\begin{align}
\ket{\Psi_0}=\prod_l \sum_{\sigma_l}u_{\sigma_l-}^l \ket{\sigma_l},
\end{align}
where $u_{\sigma_l n_l}^l(\sigma_l=\uparrow,\downarrow, n_l=+,-)$ is the matrix element of $\hat U_l'$. 
Furthermore, the expectation value of $2\hat{S}_{A_2}=\sum_l \bm e_l\cdot \hat{\bm S}_l$ due to the ground state 
is obtained as $2\langle\hat{S}_{A_2}\rangle_0=2$ obviously. 

Let us consider the following expansion of the ground state 

\begin{align}
\ket{\Psi_0}
	&=\sum_{S_{34}=0,1}\sum_{S_{12}=0,1}\sum_{S^z_{34}=-S_{34}}^{S_{34}}\sum_{S^z_{12}=-S_{12}}^{S_{12}}C_{S_{12},S^z_{12};S_{34},S^z_{34}}\nonumber\\ 
	&\hspace{35mm}\times\ket{S_{12},S^z_{12}}\ket{S_{34},S^z_{34}},
\label{gs}
\end{align}
where $S_{ij}$ and $S^z_{ij}$ are the magnitude of the total spin of two sites $(i,j)$ and 
the corresponding $z$-component, respectively. It is possible to decompose the expectation value $\langle\hat{S}_{A_2}\rangle_0$ into 
all spin-pair components, as follows
\begin{align}
&\braket{\hat S_{A_2}}_0
	=\braket{\Psi_0|\hat S_{A_2}|\Psi_0}\nonumber\\
	&=\sum_{S'_{34}=0,1}\sum_{S'_{12}=0,1}\sum_{{S^z}'_{34}=-S'_{34}}^{S'_{34}}
\sum_{{S^z}'_{12}=-S'_{12}}^{S'_{12}} \sum_{S_{34}=0,1}\sum_{S_{12}=0,1}\sum_{S^z_{34}=-S_{34}}^{S_{34}} \sum_{S^z_{12}=-S_{12}}^{S_{12}}\nonumber\\
	&\qquad\times\braket{\braket{S'_{12},{S^z}'_{12};S'_{34},{S^z}'_{34}|\hat{S}_{A_2}|S_{12},S^z_{12};S_{34},S^z_{34}}},
\end{align}
with
\begin{align}
&\braket{\braket{S'_{12},{S^z}'_{12};S'_{34},{S^z}'_{34}|
\hat{S}_{A_2}|S_{12},S^z_{12};S_{34},S^z_{34}}}\nonumber\\
 &\quad \equiv C^*_{S'_{12},{S^z}'_{12};S'_{34},{S^z}'_{34}}C_{S_{12},S^z_{12};S_{34},S^z_{34}}
 \nonumber\\
 &\qquad \times\braket{S'_{12},{S^z}'_{12};S'_{34},{S^z}'_{34}|
\hat{S}_{A_2}|S_{12},S^z_{12};S_{34},S^z_{34}},
\end{align}
by which the dominant component for the All-Out state will be detected. 
As usual, we introduce the projection operators that extract spin-singlet and spin-triplet states from individual spin states $\ket{\sigma_i\sigma_j}$ at $i$- and $j$-sites:
\begin{align}
P_{ij}^s &= -\bm S_i \cdot \bm S_j + \frac{1}{4} \quad(\rm{singlet})\\
P_{ij}^t &= \bm S_i \cdot \bm S_j + \frac{3}{4} \quad(\rm{triplet}).
\end{align}
It is obvious that these projection operators extract one spin-singlet state ($\ket{S_{ij}=0, S_{ij}^z=0}=(\ket{\uparrow\downarrow}-\ket{\downarrow\uparrow})/\sqrt 2$) and three spin-triplet states ($\ket{S_{ij}=1, S_{ij}^z=0}=(\ket{\uparrow\downarrow}+\ket{\downarrow\uparrow})/\sqrt 2, 
\ket{S_{ij}=1, S_{ij}^z=1}=\ket{\uparrow\uparrow}$ and $\ket{S_{ij}=1, S_{ij}^z=-1}=\ket{\downarrow\downarrow}$), respectively. Thus, the  transformation from the basis $\{\ket{\sigma_i\sigma_j}\}$  to the basis $\{\ket{S_{ij}S_{ij}^z}\}$ is given by the $4\times4$ unitary matrix
\begin{align}
\begin{pmatrix}
	\ket{0, 0}\\
	\ket{1, 0}\\
	\ket{1, 1}\\
	\ket{1, -1}\\
\end{pmatrix}
=\hat P_{ij}
\begin{pmatrix}
	\ket{\uparrow\downarrow}\\
	\ket{\downarrow\uparrow}\\
	\ket{\uparrow\uparrow}\\
	\ket{\downarrow\downarrow}\\
\end{pmatrix}\quad
\hat P_{ij}=
\begin{pmatrix}
	\frac{1}{\sqrt 2} & -\frac{1}{\sqrt 2} & 0 & 0 \\
	\frac{1}{\sqrt 2} & \frac{1}{\sqrt 2} & 0 & 0 \\
	0 & 0 & 1 & 0 \\
	0 & 0 & 0 & 1 \\
\end{pmatrix},
\end{align}
and the coefficient $C_{S_{12},S^z_{12};S_{34},S^z_{34}}$ in Eq.(\ref{gs}) is written, as follows
\begin{align}
&C_{S_{12},S^z_{12};S_{34},S^z_{34}}\nonumber\\
	&=\sum_{\sigma_1,\sigma_2,\sigma_3,\sigma_4}u_{\sigma_1-}^1u_{\sigma_2-}^2u_{\sigma_3-}^3u_{\sigma_4-}^4(P_{12})_{S_{12}S^z_{12},\sigma_1\sigma_2}(P_{34})_{S_{34}S^z_{34},\sigma_3\sigma_4}.
\end{align}

Among all components of $\braket{\hat S_{A_2}}_0$, the largest magnitude is obtained by 
$\braket{\braket{1,-1;1,1|\hat{S}_{A_2}|1,-1;1,1}}$=0.2234, 
while $\braket{\braket{ 0,0;0,0|\hat{S}_{A_2}|0,0;0,0}}$ vanishes (Appendix C). For simplicity, $(S'_{12},S'_{34}|\hat{S}_{A_2}|S_{12},S_{34})$ is introduced, as follows
\begin{align}
&(S'_{12},S'_{34}|\hat{S}_{A_2}|S_{12},S_{34})\nonumber\\
	&\quad\equiv\sum_{{S^z}'_{34}=-S'_{34}}^{S'_{34}}\sum_{{S^z}'_{12}=-S'_{12}}^{S'_{12}}\sum_{S^z_{34}=-S_{34}}^{S_{34}}\sum_{S^z_{12}=-S_{12}}^{S_{12}}\nonumber\\
	&\qquad\times\braket{\braket{S'_{12},{S^z}'_{12};S'_{34},{S^z}'_{34}|\hat{S}_{A_2}|S_{12},S^z_{12};S_{34},S^z_{34}}}.
\end{align}
These values $(S'_{(1,2)},S'_{(3,4)}|\hat{S}_{A_2}|S_{(1,2)},S_{(3,4)})$ are given in Table \ref{t3}. 
\begin{table}[t]
	\centering
    \caption{$(S'_{(1,2)},S'_{(3,4)}|\hat{S}_{A_2}|S_{(1,2)},S_{(3,4)})$ with 
    $a=0.0555$}\vspace{2mm}
	\begin{tabular}{c|cccc}
	\hline
	\diagbox{$(S_1,S_2)$}{$(S_3,S_4)$} & $(0,0)$ & $(0,1)$ & $(1,0)$ & $(1,1)$\\ \hline
	$(0,0)$ & 0 & $a$ & $a$ & 0\\
	$(0,1)$ & $a$ & $a$ & 0 & $2a$\\
	$(1,0)$ & $a$ & 0 & $a$ & $2a$\\
	$(1,1)$ & 0 & $2a$ & $2a$ & $4a$\\ \hline
    \end{tabular}
    \label{t3}
\end{table}
From these results, it is understood that the order parameter of the All-In (and All-Out) spin-state 
will be contributed by triplet-triplet and singlet-triplet spin states, mainly, 
while there is no contribution from the singlet-singlet spin state. 
This means that the electronic system around the All-In-All-Out spin ordered phase 
cannot take place spin-singlet superconductivity but spin-triplet superconductivity.

\section{Conclusion}
In summary, we investigate superconductivity around the AIAO spin ordered phase. The electronic system is described by the pyrochlore lattice Hubbard model with an anti-symmetric spin-orbit coupling, where there are 12 types of spin clusters. 
In particular, the AIAO spin-cluster is classified into $\rm{A_2}$ representation of $\rm{T_d}$ point group. Spin-cluster susceptibilities are calculated within the RPA with respect to the on-site electron-electron interaction. We find that AIAO spin fluctuation is predominant in all spin clusters for $\lambda<0$ around half-filling.  For the electronic system around the AIAO spin ordered phase, we have calculated the superconducting transition by solving the linearized gap equation. It is found that spin-triplet superconductivity belonging to $\rm{T_1}$ representation appears around the AIAO spin ordered phase. The reason why spin-triplet pairs instead of spin-singlet ones are favorite is examined by studying the All-Out spin state. 

Finally, we comment on the effect of the intersite Coulomb interaction term, which has been taken into account in the previous study.\cite{STI}
According to Ref. 16, we call the nearest neighbor Coulomb interaction constant 
$V$.
 Even in the electron density close to the half-filling, the AIAO spin ordered phase from the semi-metallic phase 
and a charge density wave phase will be expected by increasing $U$ and 
$V$, respectively. 
In addition, the spin ordered phase from the topological Mott-insulator (TMI) will be different from the AIAO-type. 
Furthermore, the semi-metallic phase and the TMI phase will be replaced by a metallic phase. 
Within the metallic phase, spin-triplet superconductivity will appear in the region close to the phase boundary to the AIAO phase.





\appendix
\section{Spin-cluster Operators}
The representation of spin-cluster space is reduced to $\rm T_1 \otimes \rm{\Gamma_o} =\rm A_2 \oplus \rm E \oplus 2\rm T_1 \oplus \rm T_2$, namely, 12 types of spin-clusters exist. By using the Clebsch-Gordan coefficients\cite{CG}, spin-cluster operators for these representations are described by linear combinations of the spin operators as follows,
\begin{align}
\hat S_{A_2}(\bm R)
	&= -\frac{1}{2\sqrt 3}\left(\hat S_{\bm R1}^x-\hat S_{\bm R2}^x-\hat S_{\bm R3}^x+\hat S_{\bm R4}^x-\hat S_{\bm R1}^y+\hat S_{\bm R2}^y\right.\nonumber\\
&\left.\qquad-\hat S_{\bm R3}^y+\hat S_{\bm R4}^y-\hat S_{\bm R1}^z-\hat S_{\bm R2}^z+\hat S_{\bm R3}^z+\hat S_{\bm R4}^z\right),\\
\hat S_{\rm Eu}(\bm R)
	&=-\frac{1}{2\sqrt 2}\left(-\hat S_{\bm R1}^x+\hat S_{\bm R2}^x+\hat S_{\bm R3}^x-\hat S_{\bm R4}^x\right.\nonumber\\
	&\hspace{25mm}\left.-\hat S_{\bm R1}^y+\hat S_{\bm R2}^y-\hat S_{\bm R3}^y+\hat S_{\bm R4}^y\right),\\
\hat S_{\rm Ev}(\bm R)
	&=-\frac{1}{2\sqrt 6}\left(-\hat S_{\bm R1}^x+\hat S_{\bm R2}^x+\hat S_{\bm R3}^x-\hat S_{\bm R4}^x+\hat S_{\bm R1}^y-\hat S_{\bm R2}^y\right.\nonumber\\
	&\left.+\hat S_{\bm R3}^y-\hat S_{\bm R4}^y-2\hat S_{\bm R1}^z-2\hat S_{\bm R2}^z+2\hat S_{\bm R3}^z+2\hat S_{\bm R4}^z\right),\\
\hat S_{\rm T_11x}(\bm R)
	&=\frac{1}{2}(\hat S_{\bm R1}^x+\hat S_{\bm R2}^x+\hat S_{\bm R3}^x+\hat S_{\bm R4}^x),\\
\hat S_{\rm T_11y}(\bm R)
	&=\frac{1}{2}(\hat S_{\bm R1}^y+\hat S_{\bm R2}^y+\hat S_{\bm R3}^y+\hat S_{\bm R4}^y),\\
\hat S_{\rm T_11z}(\bm R)
	&=\frac{1}{2}(\hat S_{\bm R1}^z+\hat S_{\bm R2}^z+\hat S_{\bm R3}^z+\hat S_{\bm R4}^z),\\
\hat S_{\rm T_12x}(\bm R)
	&=-\frac{1}{2\sqrt 2}\left(\hat S_{\bm R1}^y+\hat S_{\bm R2}^y-\hat S_{\bm R3}^y-\hat S_{\bm R4}^y\right.\nonumber\\
	&\hspace{25mm}\left.+\hat S_{\bm R1}^z-\hat S_{\bm R2}^z+\hat S_{\bm R3}^z-\hat S_{\bm R4}^z\right),\\
\hat S_{\rm T_12y}(\bm R)
	&=-\frac{1}{2\sqrt 2}\left(-\hat S_{\bm R1}^z+\hat S_{\bm R2}^z+\hat S_{\bm R3}^z-\hat S_{\bm R4}^z\right.\nonumber\\
	&\hspace{25mm}\left.+\hat S_{\bm R1}^x+\hat S_{\bm R2}^x-\hat S_{\bm R3}^x-\hat S_{\bm R4}^x\right),\\
\hat S_{\rm T_12z}(\bm R)
	&=-\frac{1}{2\sqrt 2}\left(\hat S_{\bm R1}^x-\hat S_{\bm R2}^x+\hat S_{\bm R3}^x-\hat S_{\bm R4}^x\right.\nonumber\\
	&\hspace{25mm}\left.-\hat S_{\bm R1}^y+\hat S_{\bm R2}^y+\hat S_{\bm R3}^y-\hat S_{\bm R4}^y\right),\\
\hat S_{\rm T_2\xi}(\bm R)
	&=-\frac{1}{2\sqrt 2}\left(\hat S_{\bm R1}^y+\hat S_{\bm R2}^y-\hat S_{\bm R3}^y-\hat S_{\bm R4}^y\right.\nonumber\\
	&\hspace{25mm}\left.-\hat S_{\bm R1}^z+\hat S_{\bm R2}^z-\hat S_{\bm R3}^z+\hat S_{\bm R4}^z\right),\\
\hat S_{\rm T_2\eta}(\bm R)
	&=-\frac{1}{2\sqrt 2}\left(-\hat S_{\bm R1}^z+\hat S_{\bm R2}^z+\hat S_{\bm R3}^z-\hat S_{\bm R4}^z\right.\nonumber\\
	&\hspace{25mm}\left.-\hat S_{\bm R1}^x-\hat S_{\bm R2}^x+\hat S_{\bm R3}^x+\hat S_{\bm R4}^x\right),\\
\hat S_{\rm T_2\zeta}
	&=-\frac{1}{2\sqrt 2}\left(\hat S_{\bm R1}^x-\hat S_{\bm R2}^x+\hat S_{\bm R3}^x-\hat S_{\bm R4}^x\right.\nonumber\\
	&\hspace{25mm}\left.+\hat S_{\bm R1}^y-\hat S_{\bm R2}^y-\hat S_{\bm R3}^y+\hat S_{\bm R4}^y\right),
\end{align}
where $\hat S_{\bm Ri}^\alpha \ (i=1,2,3,4,\mu=x, y,z)$ is $\alpha$-component of spin operatore at the $i$-th sublattice.

\section{Pairing Interaction Mediated by AIAO Spin Fluctuation}
In order to get the pairing interaction mediated by AIAO spin fluctuation, we calculate the second-order perturbation of the anomalous self-energy matrix $\hat \Delta(\bm k)$ with respect to $U$. 
According to the Dyson-Gorkov equation, the anomalous Green's function matrix is described by $\hat \Delta(\bm k)$,
\begin{align}
\hat F(\bm k, i\omega_m)
	&=-\hat G^{(0)}(\bm k, i\omega_m)\hat \Delta(\bm k, i\omega_m)\hat G(-\bm k, -i\omega_m).
\end{align}
Fig.\ref{a1} shows the diagrams of the first-order perturbation of anomalous self-energy $\Delta^{(1)}(\bm k, i\omega_m)$
\begin{align}
\Delta_{\sigma_1\sigma_2}^{(1)l_1l_2}(\bm k, i\omega_m)
	=\frac{U}{\beta N}\sum_{\bm q}\sum_m \delta_{l_1,l_2}\delta_{\bar{\sigma_1},\sigma_2}F_{\sigma_1\sigma_2}^{l_1l_2}(\bm q, i\omega_m)
\end{align}
Similarly, Fig.\ref{a2} shows diagrams of the second-order perturbation of anomalous self-energy $\hat \Delta^{(2)}(\bm k, i\omega_m)$.
Especially, the expression corresponding to Fig.\ref{a2}(a) is given as follows:
\begin{align}
&\Delta_{\sigma_1\sigma_2}^{(2a)l_1l_2}(\bm k, i\omega_m)
	=\frac{U^2}{\beta^3 N^2}\sum_{\bm q_1,\bm q_2} \sum_{m_1,m_2}G_{\sigma_2\bar{\sigma_1}}^{(0)l_2l_1}(\bm q_1,i\omega_{m_1})\nonumber\\ 
	&\times G_{\bar{\sigma_1} \sigma_2}^{(0)l_1l_2}(\bm k+\bm q_1-\bm q_2,i\omega_m+i\omega_{m_1}-i\omega_{m_2})F_{\sigma_1\bar \sigma_2}^{l_1l_2}(\bm q_2,i\omega_{m_2}),
\end{align} 

\begin{figure}[t]
\centering
\includegraphics[scale=0.55]{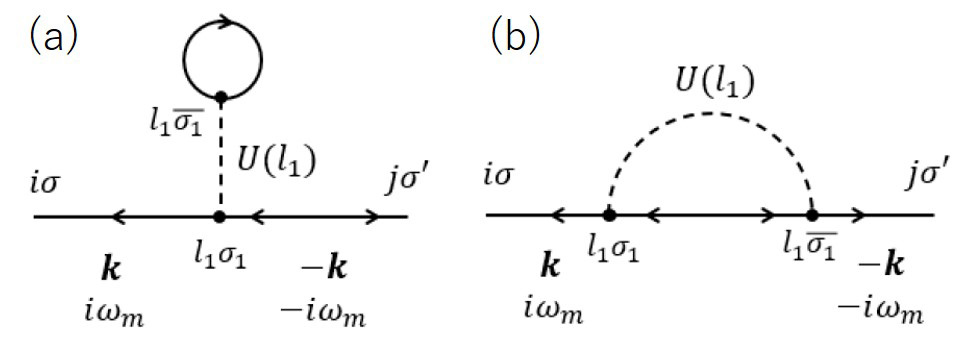}
\caption{Diagrams of the first-order perturbation of anomalous self-energy $\hat \Delta^{(1)}(\bm k, i\omega_m)$.(Color online)}
\label{a1}
\centering
\includegraphics[scale=0.48]{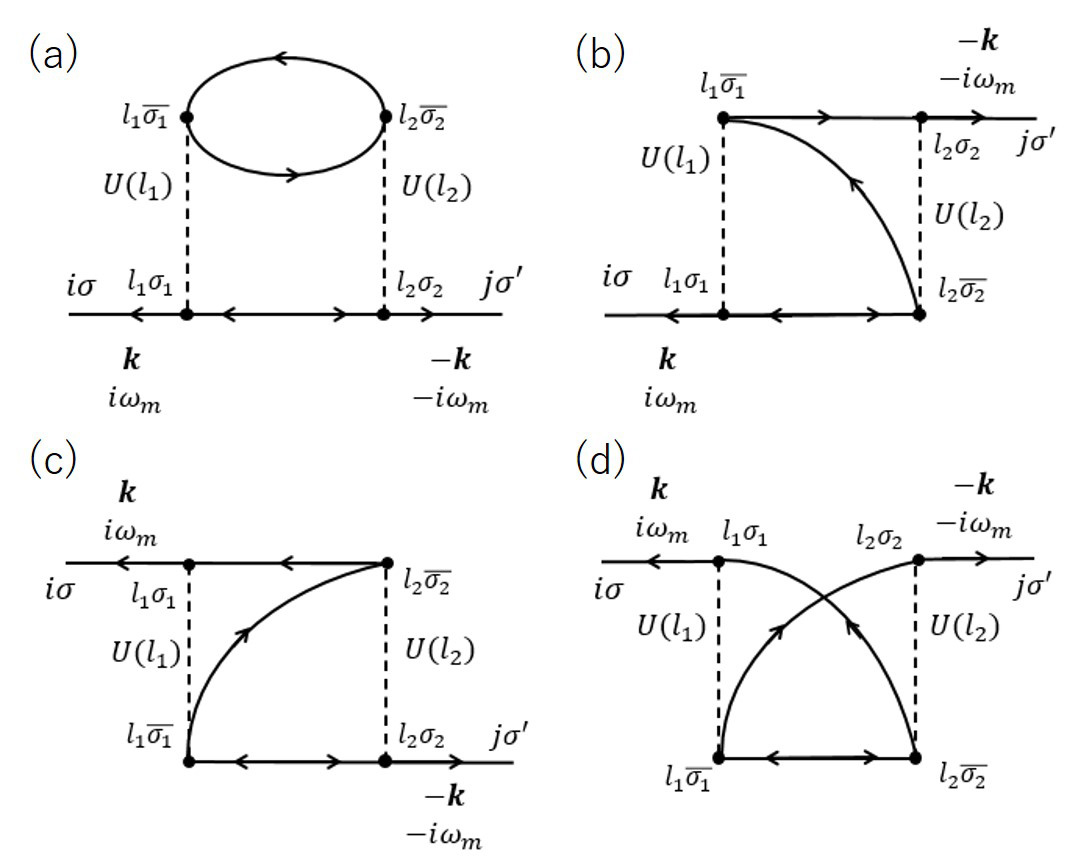}
\caption{Diagrams of the second-order perturbation of anomalous self-energy $\hat \Delta^{(2)}(\bm k, i\omega_m)$.(Color online)}
\label{a2}
\end{figure} 
Summing up four diagrams, the second-order anomalous sefl-energy $\hat \Delta^{(2)}$ ($=\hat \Delta^{(2a)}+\hat \Delta^{(2b)}+\hat \Delta^{(2c)}+\hat \Delta^{(2d)}$) is given by 
\begin{align}
&\Delta_{\sigma_1\sigma_2}^{(2)l_1l_2}(\bm k,i\omega_m)\nonumber\\
	&=-\frac{U^2}{\beta^2 N^2}\sum_{\bm q_1,\bm q_2} \sum_{m_1,m_2} \sum_{\sigma_3,\sigma_4}\sigma_1\sigma_2\sigma_3\sigma_4G_{\bar{\sigma_4}\bar{\sigma_1}}^{(0)l_2l_1}(\bm q_1,i\omega_{m_1})\nonumber\\ 
	&\quad \times G_{\bar{\sigma_3}\bar{\sigma_2}}^{(0)l_1l_2}(\bm k+\bm q_1-\bm q_2,i\omega_m+i\omega_{m_1}-i\omega_{m_2})F_{\sigma_3\sigma_4}^{l_1l_2}(\bm q_2,i\omega_{m_2})\nonumber\\
	&=-\frac{U^2}{\beta N}\sum_{\bm q} \sum_{m_2} \sum_{\sigma_3,\sigma_4}e^{-i(\bm k-\bm q)\cdot (\bm R_{l_1}-\bm R_{l_2})}\sigma_1\sigma_2\sigma_3\sigma_4\nonumber\\ 
	&\quad\times \bar {\tilde \chi}_{l_1\bar{\sigma_3}\bar{\sigma_1},l_2\bar{\sigma_4}\bar{\sigma_2}}(\bm k-\bm q,i\omega_m-i\omega_{m_2})F_{\sigma_3\sigma_4}^{l_1l_2}(\bm q,i\omega_{m_2}),
\end{align}
where $\bar {\tilde \chi}$ is the irreducible susceptibility given as
\begin{align}
&\bar {\tilde \chi}_{i\sigma_1\sigma_2, j\sigma_3\sigma_4}(\bm q,i\Omega_m)\nonumber\\
	&=-\frac{1}{\beta N}e^{i\bm q\cdot (\bm R_i-\bm R_j)}\sum_{\bm k,m'}G_{\sigma_3\sigma_1}^{(0)l_2l_1}(\bm k, i\omega_{m'})G_{\sigma_2\sigma_4}^{(0)l_1l_2}(\bm k+\bm q, i\Omega_m+i\omega_{m'}).
\end{align}

The expression of $\hat \Delta(\bm k,i\omega_m)$ calculated within RPA is yielded by replacing the noninteracting susceptibilities with the RPA ones. 
In order to use the spin cluster susceptibility, the unitary transformations due to $\hat W_1$ and $\hat W_2$ will be carried out, where $\hat W_1$ and $\hat W_2$ are shown in the main text. 
Since the AIAO spin-cluster susceptibility is significantly larger around the $\Gamma$-point than the other spin-cluster ones, $\chi_{\rm A_2}(\bm q)$ is remained only among all spin-cluster susceptibilities.
According to the weak coupling theory for superconductivity, the frequency dependence of the anomalous self-energy will be ignored. Finally, we get the pairing interaction mediated by AIAO fluctuation:
\begin{align}
&V_{\sigma_1\sigma_2\sigma_3\sigma_4}^{l_1l_2}(\bm k,\bm k')\nonumber\\
	&=\frac{U}{N}\delta_{l_1,l_2}\delta_{\sigma_2,\bar\sigma_1}\delta_{\sigma_1,\sigma_3}\delta_{\sigma_2,\sigma_4}-\frac{U^2}{N}e^{-i(\bm k-\bm k')\cdot(\bm R_{l_1}-\bm R_{l_2})}\sigma_1\sigma_2\sigma_3\sigma_4\nonumber\\
	&\times\sum_{\alpha,\alpha'}(W_{1}^{*})_{l_1\alpha,l_1\bar{\sigma_1}\bar{\sigma_3}}(W_1)_{l_2\alpha',l_2\bar{\sigma_4}\bar{\sigma_2}} (W_{2}^{*})_{{\rm A_2}, l_1\alpha}(W_{2})_{{\rm A_2},\it{l}_2\alpha'}\chi_{\rm{A_2}}(\bm k-\bm k'),
\end{align}
where $\alpha (=x,y,z)$ is the component of spin opetator.

\section{Calculation of \\$\braket{\braket{S'_{12},{S^z}'_{12};S'_{34},{S^z}'_{34}|\hat{S}_{A_2}|S_{12},S^z_{12};S_{34},S^z_{34}}}$}
The ground state $\ket{\Psi_0}$ of $H_{AO}$ 
\begin{align}
H_{AO}=-2h\:\hat{S}_{A_2}
\end{align}
can be expanded as
\begin{align}
\ket{\Psi_0}
	&=\sum_{S_{34}=0,1}\sum_{S_{12}=0,1}\sum_{S^z_{34}=-S_{34}}^{S_{34}}\sum_{S^z_{12}=-S_{12}}^{S_{12}}C_{S_{12},S^z_{12};S_{34},S^z_{34}}\nonumber\\ 
	&\hspace{35mm}\times\ket{S_{12},S^z_{12}}\ket{S_{34},S^z_{34}},
\end{align}
where $S_{ij}$ and $S^z_{ij}$ are the magnitude of the total spin of two sites $(i,j)$ and the corresponding $z$-component of the, respectively, The expectation value of $\hat S_{A_2}$ due to the ground state is obtained as $\braket{\hat S_{A_2}}_0=1$. 
Using the expansion of $\ket{\Psi_0}$ given above, the expectation value will be written as
\begin{align}
&\braket{\hat S_{A_2}}_0
	=\braket{\Psi_0|\hat S_{A_2}|\Psi_0}\nonumber\\
	&=\sum_{S'_{34}=0,1}\sum_{S'_{12}=0,1}\sum_{{S^z}'_{34}=-S'_{34}}^{S'_{34}}
\sum_{{S^z}'_{12}=-S'_{12}}^{S'_{12}} \sum_{S_{34}=0,1}\sum_{S_{12}=0,1}\sum_{S^z_{34}=-S_{34}}^{S_{34}} \sum_{S^z_{12}=-S_{12}}^{S_{12}}\nonumber\\
	&\qquad\times\braket{\braket{S'_{12},{S^z}'_{12};S'_{34},{S^z}'_{34}|\hat{S}_{A_2}|S_{12},S^z_{12};S_{34},S^z_{34}}},
\end{align}
with
\begin{align}
&\braket{\braket{ S'_{12},{S^z}'_{12};S'_{34},{S^z}'_{34}|\hat{S}_{A_2}|S_{12},S^z_{12};S_{34},S^z_{34}}}\nonumber\\
	&\qquad \equiv C^*_{S'_{12},{S^z}'_{12};S'_{34},{S^z}'_{34}}C_{S_{12},S^z_{12};S_{34},S^z_{34}}
 \nonumber\\
	&\qquad \times \braket{S'_{12},{S^z}'_{12};S'_{34},{S^z}'_{34}|\hat{S}_{A_2}|S_{12},S^z_{12};S_{34},S^z_{34}}.
\end{align}
The component of the singlet-singlet spin state is zero:
\begin{align}
\braket{\braket{0,0;0,0|\hat{S}_{A_2}|0,0;0,0}}=0.
\end{align}
The non-zero components of the singlet-triplet spin states are evaluated as follows.
\begin{align}
&\braket{\braket{0,0;0,0|\hat{S}_{A_2}|0,0;1,1}}=0.043815,\nonumber\\
&\braket{\braket{0,0;0,0|\hat{S}_{A_2}|0,0;1,-1}}=0.01174,\nonumber\\
&\braket{\braket{0,0;0,0|\hat{S}_{A_2}|1,1;0,0}}=0.01174,\nonumber\\
&\braket{\braket{0,0;0,0|\hat{S}_{A_2}|0,0;1,-1}}=0.043815,\nonumber\\
&\braket{\braket{0,0;1,1|\hat{S}_{A_2}|0,0;1,1}}=0.059853,\nonumber\\
&\braket{\braket{ 0,0;1,1|\hat{S}_{A_2}|0,0;1,-1}}=0.004297,\nonumber\\
&\braket{\braket{0,0;1,1|\hat{S}_{A_2}|1,1;1,1}}=0.021908,\nonumber\\
&\braket{\braket{0,0;1,-1|\hat{S}_{A_2}|1,1;1,-1}}=0.001573,\nonumber\\
&\braket{\braket{0,0;1,1|\hat{S}_{A_2}|1,-1;1,1}}=0.08176,\nonumber\\
&\braket{\braket{0,0;1,-1|\hat{S}_{A_2}|1,-1;1,-1}}=0.00587,\nonumber\\
&\braket{\braket{1,1;0,0|\hat{S}_{A_2}|1,1;0,0}}=-0.004297,\nonumber\\
&\braket{\braket{1,-1;0,0|\hat{S}_{A_2}|1,-1;0,0}}=0.059853,\nonumber\\
&\braket{\braket{1,1;0,0|\hat{S}_{A_2}|1,1;1,1}}=0.00587,\nonumber\\
&\braket{\braket{1,1;0,0|\hat{S}_{A_2}|1,1;1,-1}}=0.001573,\nonumber\\
&\braket{\braket{1,-1;0,0|\hat{S}_{A_2}|1,-1;1,1}}=0.08176,\nonumber\\
&\braket{\braket{1,-1;0,0|\hat{S}_{A_2}|1,-1;1,-1}}=0.021908.\nonumber
\end{align}
The non-zero components of the triplet-triplet spin states are evaluated as follows,
\begin{align}
&\braket{\braket{1,1;1,-1|\hat{S}_{A_2}|1,1;1,-1}}=-0.001151,\nonumber\\
&\braket{\braket{1,-1;1,1|\hat{S}_{A_2}|1,-1;1,1}}=0.223374.\nonumber
\end{align}
All spin-pairs components of $\hat S_{\rm A_2}$ have the symmetric property 
\begin{align}
&\braket{\braket{S'_{12},{S^z}'_{12};S'_{34},{S^z}'_{34}|\hat{S}_{A_2}|S_{12},S^z_{12};S_{34},S^z_{34}}}
\nonumber\\
	&\qquad=\braket{\braket{S_{12},S^z_{12};S_{34},S^z_{34}|\hat{S}_{A_2}|S'_{12},{S^z}'_{12};S'_{34},{S^z}'_{34}}}.
\end{align}

\end{document}